\documentclass[intlimits,twoside,a4paper]{article}

\usepackage{amsmath,amssymb}
\usepackage{graphicx}
\usepackage{color}

\usepackage[T2A]{fontenc}
\usepackage[cp1251]{inputenc}

\usepackage[eqsecnum]{cmpj2}

\issue{2017}{20}{1}{13004}
\doinumber{10.5488/CMP.20.13004}

\title[Scaling laws for random walks in long-range correlated disordered media]%
{Scaling laws for random walks in long-range correlated disordered media%
}
\author[N. Fricke \textsl{et al.}]{
  N. Fricke\refaddr{label1},
  J. Zierenberg\refaddr{label1,label2,label3},
  M. Marenz\refaddr{label1},
  F.P. Spitzner\refaddr{label1},
  V. Blavatska\refaddr{label4},
  W. Janke\refaddr{label1}
}
\addresses{
\addr{label1} Institut f\"ur Theoretische Physik,  Universit\"at Leipzig, Postfach 100\,920, D--04009 Leipzig, Germany
\addr{label2} Bernstein Center for Computational Neuroscience, Am Fassberg 17, D-37077 G{\"o}ttingen, Germany
\addr{label3} Max Planck Institute for Dynamics and Self-Organsization, Am Fassberg 17, D-37077 G{\"o}ttingen, Germany
\addr{label4} Institute for Condensed Matter Physics
of the National Academy of Sciences of Ukraine,\\
1 Svientsitskii St., 79011 Lviv, Ukraine
}
\authorcopyright{N. Fricke, J. Zierenberg, M. Marenz, F.P. Spitzner, V. Blavatska, W. Janke, 2017}
\date{Received January 27, 2017, in final form March 14, 2017}

\DeclareMathOperator{\erfc}{erfc}

\begin{document}

\maketitle

\begin{abstract}
	We study the scaling laws of diffusion in two-dimensional media with long-range
	correlated disorder through exact enumeration of random walks.  The
  disordered medium is modelled by percolation clusters with correlations
  decaying with the distance as a power law, $r^{-a}$, generated with the
  improved Fourier filtering method. To characterize this type of disorder, we
  determine the percolation threshold $p_{\text c}$ by investigating cluster-wrapping
  probabilities. At $p_{\text c}$, we estimate the (sub-diffusive) walk dimension $d_{\text w}$
  for different correlation exponents $a$. Above $p_{\text c}$, our results suggest a
  normal random walk behavior for weak correlations, whereas anomalous diffusion
  cannot be ruled out in the strongly correlated case, i.e., for small $a$.
\pacs 05.70.Jk, 64.60.al, 64.60.De
\end{abstract}

\section{Introduction}

Structural disorder has a strong effect on a wide range of physical
processes and can alter the behavior of systems such as magnets or
polymers~\cite{Avnir1983,Avnir1984}.
Examples for disordered systems are porous materials, which often have
fractal structure. This may lead to anomalous diffusive
transport~\cite{Bouchaud1990,Malek2001}, an aspect that is relevant to problems ranging from
oil recovery through porous rocks~\cite{Dullien1979,Sahimi1995}
and the dynamics of fluids in disordered media~\cite{Skinner2013, Spanner2016}
to transport
processes in crowded biological cells~\cite{Bancaud2012, Hoefling-Franosch2013}.

Disordered systems are
conveniently described in the framework of lattice models with randomly
positioned defects.
Already a small amount of defects can drive the
critical behavior of magnetic systems into a new ``disordered''
universality class \cite{Harris1974}, but the scaling properties of diffusion
(usually modelled by random walks) and polymers (usually modelled by
self-avoiding random walks) are thought to be unaffected in this case~\cite{Harris1983, Meir-Harris1989}.
For a larger amount of defects,
near the percolation threshold of the non-defect sites,
clusters of connected sites
become fractal, i.e., self-similar objects devoid of a characteristic length scale. The case where the defects are uncorrelated
is a classic textbook model, whose properties have been studied
extensively~\cite{Stauffer1992}.
Right at the percolation point, even diffusion and
polymer statistics show a modified, anomalous behavior, which has
been extensively studied using field-theoretic
renormalization group methods~\cite{Blavatska2005,Holovatch2006} and computer
simulations~\cite{Janke2012}. For recent examples see \cite{Ferber2004, Janssen2007, Blavatska2008, Blavatska2008a, Blavatska2009, Blavatska2010, Janssen2012, Blavatska2014} and further references therein.

In nature, however, inhomogeneities are frequently not distributed totally at
random but tend to be correlated over large distances. To understand the impact
of this, it is useful to consider the limiting case where correlations decay
asymptotically  as a power law (rather than exponentially) with distance $r$:
\begin{equation}
C(r)\sim r^{-a} \label{corr}.
\end{equation}
If the correlation exponent $a$ is smaller than the spatial dimension $d$, the
correlations are considered as long-range or ``infinite''.
This problem has first been investigated in the context of spin systems and
later on for percolation~\cite{Weinrib1983,Weinrib1984}. The relevance of the
disorder was shown to be characterized by an extension of the Harris criterion:
the critical behavior of the system changes if the minimum of $d$ and $a$ is
smaller than $2/\nu$, where $\nu$ denotes the correlation-length exponent for
a pure system. Furthermore, it was argued that in the regime of long-range
correlations,
the critical correlation-length exponent for strong
disorder is always given by $2/a$.
This result is still slightly controversial~\cite{Prudnikov1999,Prudnikov1999a,Prudnikov2000},
but it has, to some extent, been supported by numerical
investigations~\cite{Prakash1992,Schrenk2013}. These studies made use of the
Fourier filtering method (FFM)~\cite{Peng1991,Prakash1992,Makse1995,Makse1996}
to generate power-law correlated disorder and have yielded estimates for the
values of various critical exponents and fractal dimensions characterizing the
disordered media.

Systems confined to media with long-range correlated disorder
show a very interesting behavior, but still comparatively few properties are fully understood to date.
The investigated examples include magnetic materials~\cite{Ballesteros1999, Blavatska2003, Blavatska2005b, Blavatska2006, Ivaneyko2008, Dudka2016} and
macromolecular systems~\cite{Blavatska2001, Blavatska2001b, Blavatska2002}, again mainly using
field-theoretic
renormalization group methods~\cite{Blavatska2005, Holovatch2006} and
computer simulations~\cite{Janke2012}.
Long-range correlations have also been shown to have an impact on the conductance of
critical clusters and on their properties as a medium for diffusion. However,
these investigations have so far focused on the system at criticality, while the
situation with fewer defects has largely been disregarded. This regime is of
little interest for the uncorrelated case, where the behavior should be the same as in the
undiluted system. However, large fluctuations caused by the
long-range correlations may have a non-trivial impact on the system's dynamic
exponents and give rise to anomalous diffusion.
In the context of continuum percolation, fluid
	dynamics and diffusive transport in correlated disordered media
have recently been investigated by simulations and experiment~\cite{Skinner2013,
Spanner2016},
and correlated random potentials could be realized experimentally with laser speckle
patterns~\cite{Bewerunge2016}.

The remainder of the article is organized as follows: in section~\ref{sect_LCD}, we
closely follow the procedure of \cite{Makse1996} and use the (improved) FFM in order
to first generate continuous variables with varying degree of correlation $a$,
which are then mapped to correlated discrete values as illustrated in
figure~\ref{figDisorderMapping}. In section~\ref{sect_pc}, we determine the
site-percolation threshold $p_{\text c}$ for a square lattice as a function of $a$. In
section~\ref{sect_RWs}, we use exact enumerations of random walks
in the thus generated long-range correlated disordered medium to study diffusion
and analyze the scaling behavior for the quenched disorder
average of the mean square displacement. This is
done for
varying degrees of correlation controlled by the exponent $a$ and
different concentrations $p$.
The conclusions of our
work and prospects for future investigations are contained in
section~\ref{sect_summary}.

%\vspace{-3mm}
\section{Long-range correlated percolation clusters}
%\vspace{-1mm}
\subsection{Generation of power-law correlations}\label{sect_LCD}

We consider a two-dimensional square lattice with $L\times L$ sites labeled by
$\vec{x}$. The goal is to obtain discrete binary site
values, \mbox{$s_{\vec{x}}\in\{0,1\}$}, that exhibit a power-law correlation.
This can be achieved in a two-step process. First, one uses the modified Fourier
filtering method (FFM)~\cite{Makse1995,Makse1996} to generate continuous
Gaussian site variables $\varphi_{\vec{x}}$ that are power-law correlated, i.e.,
\begin{equation}\label{eqCorrelationGoal}
    \langle\varphi_{\vec{x}}\varphi_{\vec{x}+\vec{r}}\rangle\sim r^{-a},
\end{equation}
for sufficiently large distances $r$. In the second step, the continuous variables
$\varphi_{\vec{x}}$ are mapped to discrete values $\{0,1\}$.

\begin{figure}[!t]
\centering \includegraphics[width=0.38\textwidth, trim=3cm 1cm 2.2cm 1cm, clip=true]{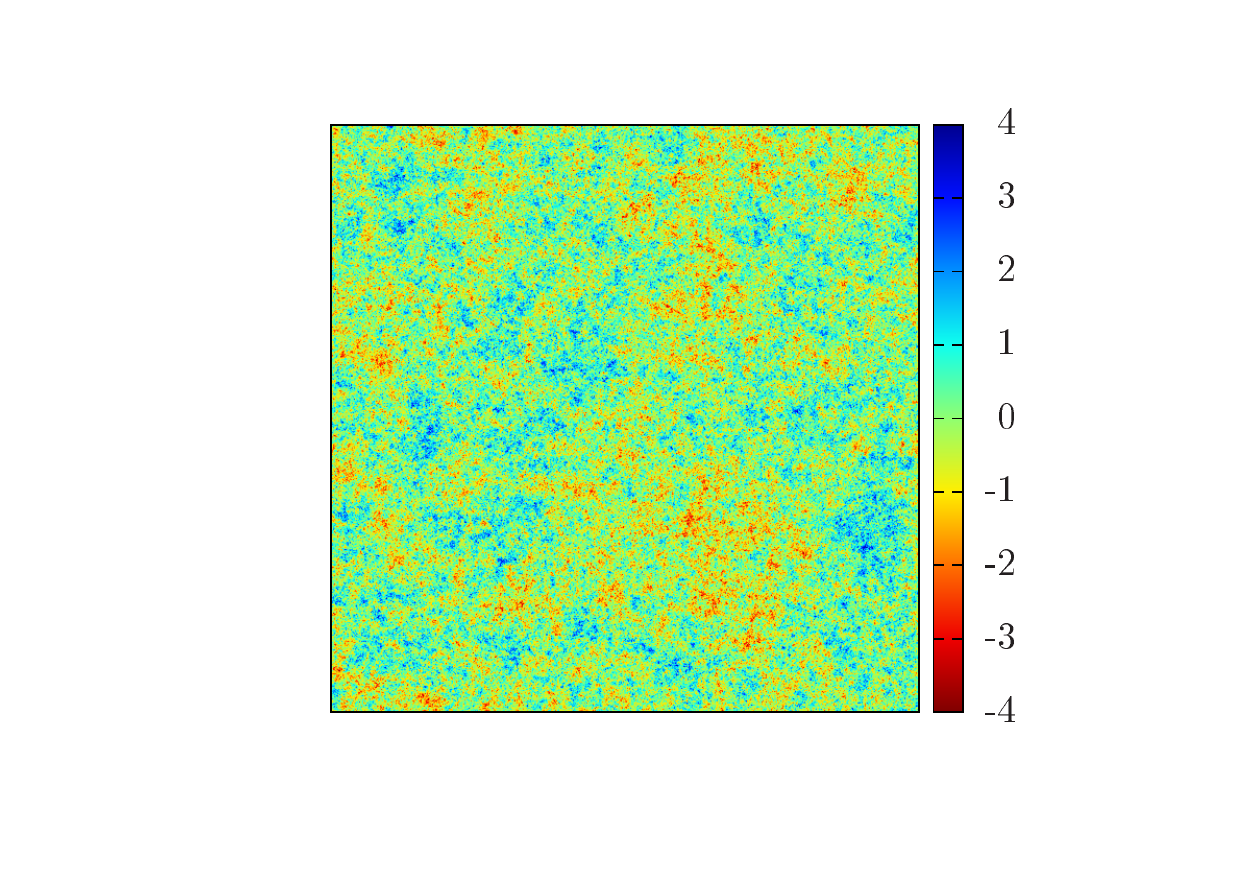}\hspace*{10mm}
\centering \includegraphics[width=0.33\textwidth,trim=3cm 1cm 3.2cm 1cm, clip=true]{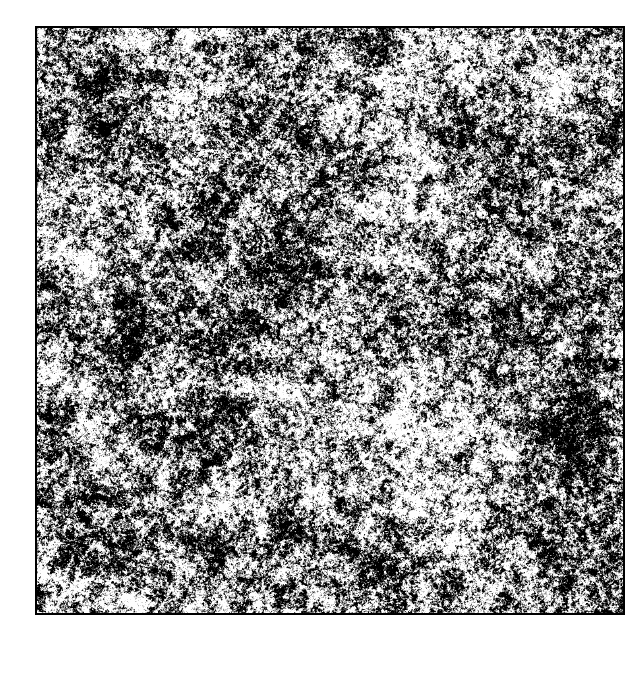}
  \vspace{-3mm}
\caption{(Color online) Correlated continuous variables on a $2^{11}\times2^{11}$
lattice for a correlation exponent $a=0.5$ (left-hand) and corresponding discrete binary variables
at the percolation threshold ($p=0.5209$) with defects shown in black (right-hand).
}\label{figDisorderMapping}
\end{figure}

The method starts with uncorrelated random numbers $u_{\vec{x}}$ drawn from a
Gaussian distribution with variance one and mean zero, which are distributed onto
the two-dimensional lattice. The key idea now is to multiply the Fourier
coefficients $u_{\vec{q}}$ with the square root of the spectral density
$S_{\vec{q}}$, which is the Fourier transform of the desired correlation
function:
\begin{equation}\label{S_u_to_phi}
  \varphi_{\vec{q}} = \sqrt{S_{\vec{q}}} \, u_{\vec{q}}\,.
\end{equation}
The inverse Fourier transform should then yield coefficients $\varphi_{\vec{x}}$
with the desired power-law correlations. The density distribution of the values,
$P(\varphi_{\vec{x}})$, is again Gaussian in the asymptotic limit of large
systems and/or many disorder realizations~\cite{Makse1995}.  The coefficients
$u_{\vec{q}}$ and $\varphi_{\vec{x}}$ can be efficiently calculated using the
discrete fast Fourier transform (FFT) algorithm~\cite{NumericalRecipiesFFT}.

\begin{figure}[!t]
\centering \includegraphics[width=0.49\textwidth]{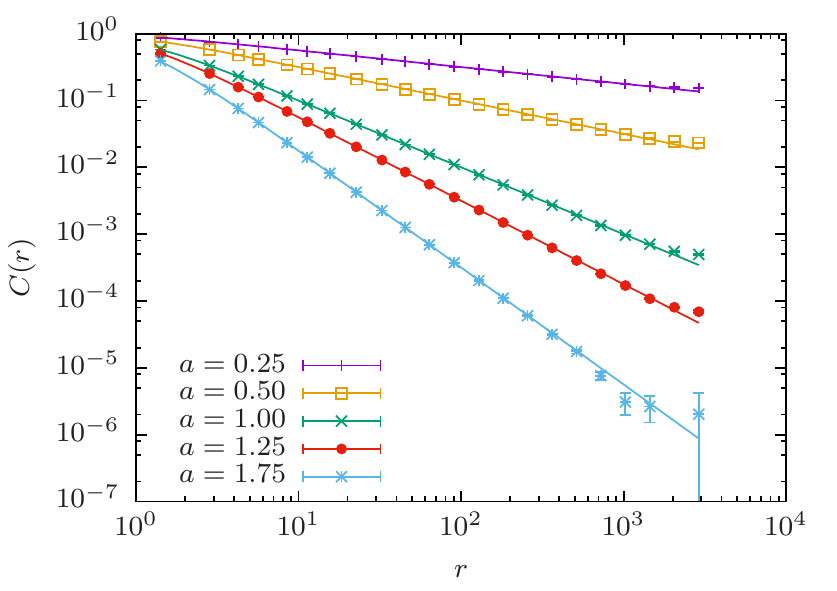}
\centering \includegraphics[width=0.49\textwidth]{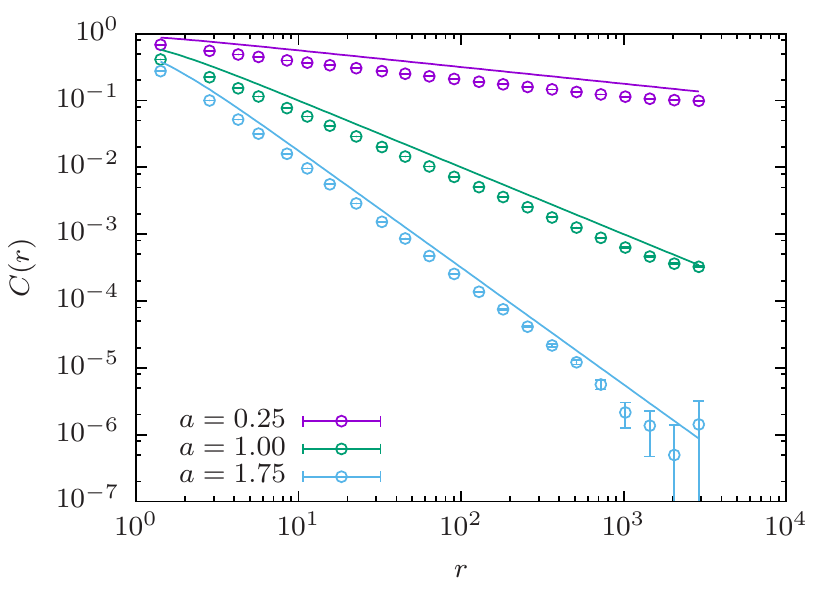}
  \caption{\label{figCorrelationSequence} (Color online) Left-hand: measured diagonal correlation function
    $\overline{C}(r)$ (data points) compared to the input correlation function
    \eqref{eqCorrelationMod} (lines) of power-law correlated continuous Gaussian
    variables on a $2^{12}\times2^{12}$ square lattice with periodic boundary conditions.
    Right-hand: the resulting diagonal correlation function of the mapped discrete disorder at
    the average percolation threshold $p=p_{\text c}$ using equation~\eqref{eq_mapping}.
    The lines show equation~\eqref{eqCorrelationMod} normalized to the expected
    variance $p(1-p)$ to serve as a guide to the eye for the asymptotic
    behavior.
    All data are obtained as quenched average from $10^5$ replica.
  }
\end{figure}

The irrelevant singularity of the correlation function \eqref{eqCorrelationGoal}
at $r=0$ poses technical problems. These can be overcome by introducing a
function with the same asymptotic large-distance limit:
\begin{equation}\label{eqCorrelationMod}
	C(r) = \left(1+r^2\right)^{-a/2} \stackrel{r \gg 1}{\longrightarrow} r^{-a}.
\end{equation}
Using a continuum approximation, the spectral density can then be calculated
analytically~\cite{Makse1996}. In $d=2$ dimensions, the result is
\begin{equation}\label{eqCorrelationS2}
  S_{\vec{q}} =
  \frac{2\pi}{\Gamma\left(\frac{a}{2}\right)}\left(\frac{|\vec{q}|}{2}\right)^{\frac{a}{2}-1}K_{\frac{a}{2}-1}\left(|\vec{q}|\right),
\end{equation}
where
$K_{\beta}(q)$ is the modified Bessel function of the order $\beta$.
\newcommand{\mzero}{|\vec{m}_0|}
This is inserted in \eqref{S_u_to_phi}, and a discrete Fourier transform
with $\vec{q}=\frac{2\pi}{L}\vec{m}$ is used, where
$-\frac{L}{2}<m_{x/y}\leqslant\frac{L}{2}$ (assuming periodic boundary conditions). To cope with the singularity of $S_{\vec{q}}$ at $q=0$ in the {\em
discrete\/} Fourier transform (which in the continuum formulation is
integrable), one assigns a suitable value $\mzero\in(0,1)$ to the zero-frequency
mode \cite{Makse1996}. The choice does not
affect the asymptotic scaling behavior but may still introduce deviations from
the desired form, especially for strong correlations.  For all cases
considered, we adjusted $\mzero$ until the empirical mean diagonal correlation
function normalized by the variance $\sigma^2$ of the distribution of random Gaussian
correlated variables,
\begin{equation}
	\overline{C}(r) = \frac{1}{2L^2\sigma^2}\left[\sum_{\vec{x},\pm}\left(\varphi_{\vec{x}}-\bar{\varphi}\right)\left(\varphi_{\vec{x}+\vec{r}_{\pm}}-\bar{\varphi}\right)\right]_{\text{av}},
\end{equation}
was found in close agreement with
equation~\eqref{eqCorrelationMod}.  Here, $\vec{r}_{\pm}=(r,\pm r)/\sqrt{2}$ and the
square bracket $[\dots]_{\text{av}}$ denotes the quenched disorder average. The mean
value $\bar{\varphi}$ of the distribution $P(\varphi)$ is zero due to symmetry.
An example of the resulting $\varphi_{\vec{x}}$ distribution is shown in the
left-hand panel of figure~\ref{figDisorderMapping} and $\overline{C}(r)$ is compared to
the target function $C(r)$ of equation~\eqref{eqCorrelationMod} in the left-hand panel of
figure~\ref{figCorrelationSequence}.  The curves slightly level off for large $r$ due
to the periodic boundary conditions.

At first glance, one might assume that the variance is $\sigma^2=1$
since the Fourier transform of the random Gaussian variables is scaled with the
structure factor according to equation~\eqref{S_u_to_phi} and
$\sigma^2=C(0)=\int_{-\infty}^{\infty} \rd\vec{q} S_{\vec{q}}=1$. However, this is
only valid for the continuous Fourier transform in infinite space, which was
considered for the analytical derivation of equation~\eqref{eqCorrelationS2}. In the FFM, we use this
as an approximation and work on a finite lattice, discretizing $S_{\vec{q}}$ in
the interval $q_{x/y}\in (-\pi,\pi]$. The lattice spacing is fixed and causes
aliasing effects, while in the infinite-size limit, the resolution of
$S_{\vec{q}}$ approaches the analytic result. This enables us to estimate the
infinite-size variance as $\sigma^2(a)=\int_{-\pi}^{\pi}
\rd{\vec{q}}S_{\vec{q}}\leqslant 1$, where unity is approached for $a\rightarrow 0$.

In the second step, the thus generated $\varphi_{\vec{x}}$ are mapped to
correlated discrete binary values, $s_{\vec{x}} \in \{0,1\}$, with a mean
density  $p$ of non-defect sites.
To this end, a threshold $\theta$ is introduced such that sites
with $\varphi_{\vec{x}}>\theta$ are considered as defects. The threshold is tied
to $p$ via
\begin{equation}\label{eq_mapping}
p(\theta)=\int_{-\infty}^{\theta}P(\varphi)\rd\varphi=\frac{1}{2}\erfc\left(\frac{-\theta}{\sqrt{2\sigma^2(a)}}\right),
\end{equation}
where $\erfc$ is the
complementary error function.
Note that the fraction of defects on individual lattices significantly fluctuates
for strong correlations and that
here we consider the quenched
disorder average $p_{L} = [p(\theta)]_{\text{av}}$ over all lattice realizations. The resulting pattern of the
power-law correlated discrete binary variables $s_{\vec{x}}$ is shown in the right-hand
panel of figure~\ref{figDisorderMapping}.
As can be seen in the right-hand panel of figure~\ref{figCorrelationSequence}, the
correlations of the lattices mapped via equation~\eqref{eq_mapping} decay with the
desired correlation exponent $a$ over a long range, though the amplitudes are
somewhat diminished.

\subsection{Percolation threshold}\label{sect_pc}

At low concentrations, clusters of the connected non-defect sites are of a typical
size, despite the power-law correlations in the system.  As for normal
percolation, we can hence define a cluster correlation length~$\xi$ that
diverges as we approach the critical concentration $p_{\text c}$. The universal exponent
$\nu_{a}$ describing this divergence is supposed to be the same as for
uncorrelated percolation, $\nu = 4/3$ in two dimensions, as long as the
correlations are sufficiently weak ($a \geqslant 2/\nu$)
according to the extended Harris criterion~\cite{Weinrib1984},
whereas for stronger correlations it is inversely proportional
to $a$~\cite{Weinrib1983,Weinrib1984}:
\begin{equation} \label{Harris}
	\nu_{a} = \left\{\begin{array}{lll}
			 \nu = 4/3 & \text{for} & a \geqslant 3/2, \\
			 2/a & \text{for}      & a <  3/2,\\
		         \end{array}
		 \right. \qquad (d=2).
\end{equation}

The value of $p_{\text c}$ is not universal but depends on the lattice type and
in general will also be affected by the presence of correlations (exceptions
are bond percolation on the square lattice and site percolation on the
triangular lattice, where $p_{\text c}$ must remain $1/2$ for
symmetry reasons~\cite{Schrenk2013}). Besides the asymptotic decay rate (i.e., the
exponent $a$), short-range aspects of the correlation function and
hence the method used to generate it should also play a role.

\begin{figure*}[!t]
	\includegraphics[width=0.49\textwidth]{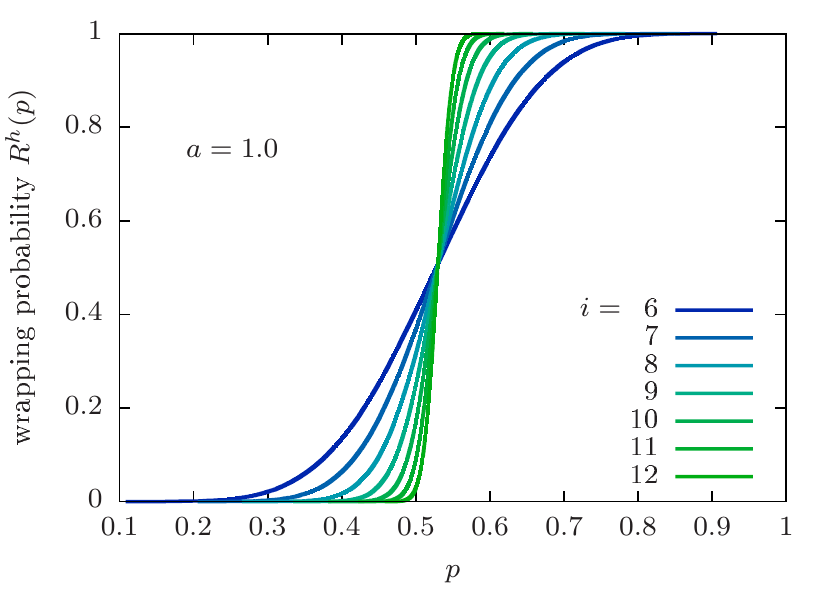}\hfill
	\includegraphics[width=0.49\textwidth]{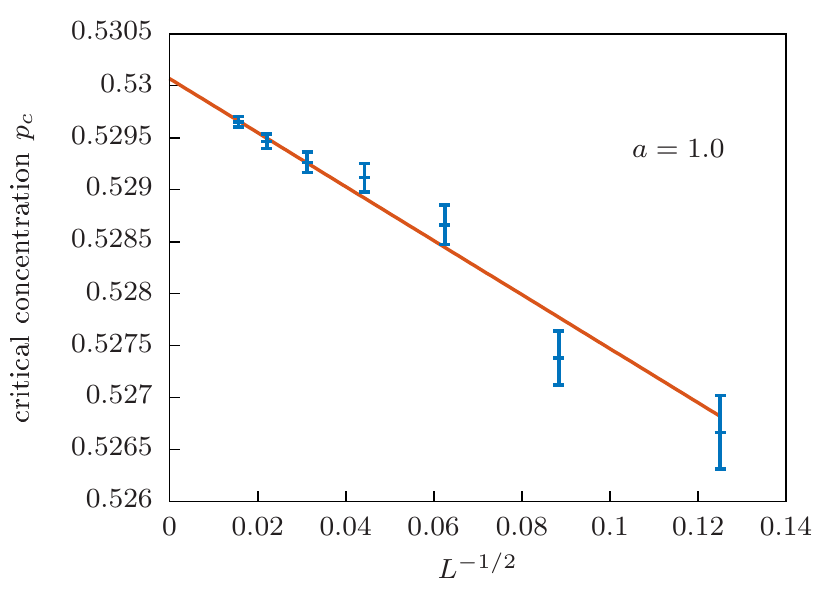}
  \caption{(Color online)
    Example for finite-size scaling of the percolation threshold for $a=1$ and
    lattice sizes $L=2^i$, $i=6,\dots,12$.
    Horizontal wrapping probabilities (left-hand) and
    critical concentrations (right-hand)
    obtained from the
    quenched disorder average of the occupation numbers, $[p(\theta)]_{\text{av}}$.
    \label{figFSSpc}
  }
\end{figure*}

To estimate $p_{\text c}(a)$, we analyzed at what concentration percolating clusters
emerge for the first time. Following \cite{Newman2001}, we used the
(horizontal) ``wrapping criterion'', which has the benefit of relatively small
finite-size effects. Accordingly, a cluster is defined as \emph{percolating} if
it closes back on itself across one specific boundary (e.g., horizontally).  The
threshold concentration was measured for a sample of $10^5$ disorder
configurations for each lattice size $L=2^i$, $i=6, \dots, 12$ for correlations
between $a=0.125$ and $a=3$. \linebreak
To do this, we adjusted the threshold $\theta$ following a simple bisection
protocol until the sites with $\varphi\leqslant\theta$  just barely percolate.
The
result was then translated back to a concentration value using
equation~\eqref{eq_mapping}
and taking the quenched disorder average,
$p_{L} = [p(\theta)]_{\text{av}}$.
We thus obtained the horizontal wrapping probabilities
$R^{\text h}(p)$ as the accumulated distributions shown in figure~\ref{figFSSpc} (left-hand).
With increasing size they become steeper, approaching a step function in the
asymptotic limit.  For stronger correlations, they are markedly more stretched
because fluctuations in the system are more pronounced. The location of the
intersection points on the $p$-axis appears to be a good estimator for $p_{\text c}$
(cf.\ \cite{Newman2001}, figure~9).

\begin{figure*}[!b]
  \centering
\includegraphics[width=0.51\textwidth]{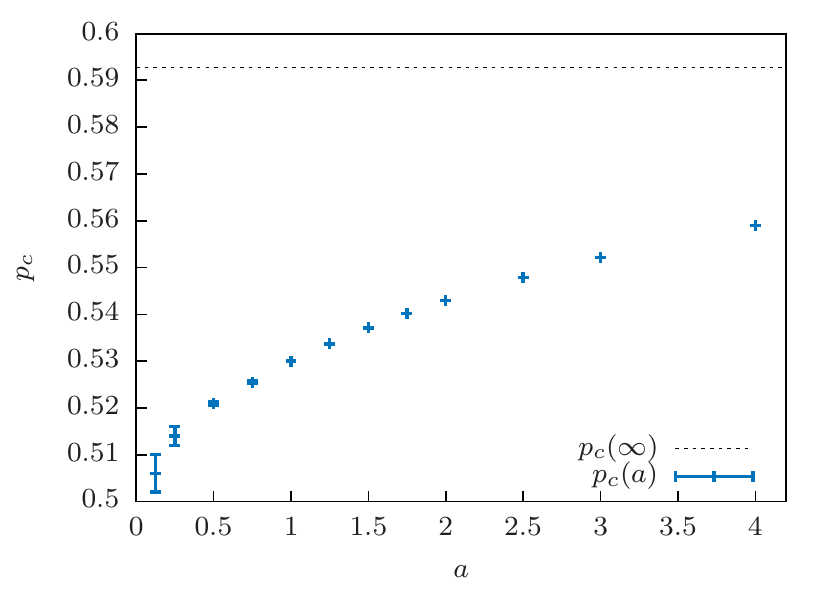}
  \caption{(Color online)
    Results of the percolation threshold  $p_{\text c}$ from the finite-size scaling
    extrapolation as a function of the correlation strength for the square
    lattice. The dashed line indicates the value for the uncorrelated case which
    is recovered for $a\rightarrow\infty$.
    \label{figPca}
  }
\end{figure*}

To obtain a more controlled estimate, however, we used a standard finite-size
scaling approach~\cite{Stauffer1992} for average threshold concentrations
$p_{L,\text c}$:
\begin{equation}
  p_{L,\text c}-p_{\text c} \sim L^{-1/\nu_a},
\end{equation}
where $\nu_a$ is given by equation~(\ref{Harris}).  Plotting $p_{L,\text c}$ vs.
$L^{-1/\nu_a}$, we hence obtain the percolation threshold $p_{\text c}$ as the intersection of the best fit with
the vertical axis as shown to the right in figure~\ref{figFSSpc} for $a=1$ where
$\nu_a = 2/a = 2$. The obtained results for $p_{\text c}$ over a wide range of correlation
strengths $a$ are plotted in figure~\ref{figPca} and listed below in
table~\ref{tab_exponents}. With increasing correlations (small $a$) the values for $p_{\text c}$
tend towards $1/2$ as was observed in a previous study~\cite{Prakash1992}.
With decreasing correlations (large $a$),
the value for uncorrelated disorder marked by the dashed line in
figure~\ref{figPca} is approached surprisingly slowly, which shows how much the value
of $p_{\text c}$ is influenced by local (short range) properties of the system.
As a check of our analysis method, we also looked at the
completely uncorrelated case and found a good
agreement with the known estimates~\cite{Ziff1994, Newman2001}. Interestingly, the
values of the critical
horizontal wrapping probabilities were all found very close to the (exactly
known~\cite{Pinson1994}) uncorrelated value $R^{\text h}_{\text c}\approx 0.52$.

\section{Random walk and diffusion exponents}\label{sect_RWs}

We now turn to the problem of diffusive random walks (RWs) in an environment with long-range correlated disorder.
In the asymptotic limit, the mean square
displacement as a function of the number of steps~$N$ follows a power law:
\begin{equation}
\left[\left \langle R^2 \right \rangle \right]_{\text{av}}\sim N^{2\nu_{\text{RW}}},
\end{equation}
with $\nu_{\text{RW}}=1/2$ in any non-fractal medium.
In fractal
systems, however, random walks are usually subdiffusive with $\nu_{\text{RW}}$ assuming
a smaller, non-trivial value, and one defines the so-called walk (fractal)
dimension
as $d_{\text w}=1/\nu_{\text{RW}}$.
For two-dimensional uncorrelated critical percolation clusters,
$d_{\text w}$ has been measured very efficiently using the Lobb-Frank
algorithm~\cite{Frank1988}, yielding a very accurate estimate of
\mbox{$d_{\text w}=2.8784(8)$}~\cite{Grassberger1999}. For any higher
concentrations, the value of $d_{\text w}$ is simply $2$ since the system becomes
homogeneous at large scales. In the presence of long-range correlated disorder,
by contrast, the spatial distribution of defects is inhomogeneous on any length
scale. This might affect the value of $d_{\text w}$.

We used a simple exact enumeration method~\cite{Majid1984a} to generate
the probability densities of RWs. Specifically, we considered
the ``blind ant'' rule where at each step,
a diffusing particle (ant)
randomly chooses its next position uniformly among all nearest neighbors. If it
picks a defect, it will bounce back to its original position. The enumeration
algorithm proceeds by successively calculating the probability distributions
$P(\vec{x}_i,N)$ in order to find the walker at any site $\vec{x}_i$ after the $N$th
step from those after the ($N-1$)th step:
 \begin{equation}\label{rw_enum}
   P(\vec{x}_i,N)= \frac{1}{4}\sum_{n} P(\vec{x}_n,N-1)+\frac{k}{4} P(\vec{x}_i,N-1),
 \end{equation}
where the sum goes over all neighboring sites and $k$ is the number of adjacent
defects.  This algorithm has polynomial complexity
$\propto N^{d_l+1}$, where $d_l$ ($\approx 1.68$ for two-dimensional
uncorrelated percolation) is the chemical dimension,
so that very long walks can
in principle be generated. From the knowledge of $P(\vec{x}_i,N)$,
all moments or cumulants of the walk displacement, the return probability, etc.
can be readily computed. However, to take a large disorder average
(here, over $10^5$ replica) can still become very time- and memory-consuming, and we, therefore,
limited our simulations to a maximum of $N=4096$ steps and focused
in the analysis on the mean square displacement of the random walks.

\begin{figure}[!ht]
  \centering
 \includegraphics[width=0.43\textwidth]{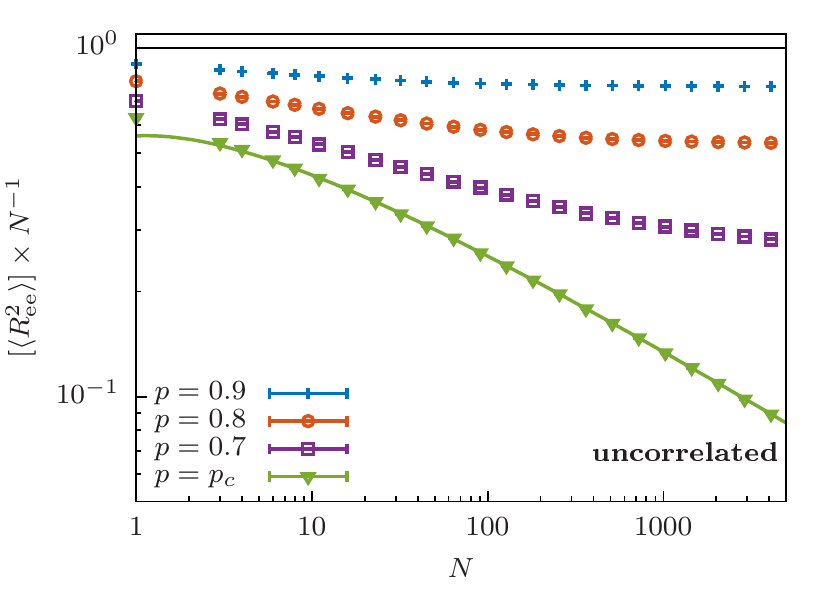}
  \includegraphics[width=0.43\textwidth]{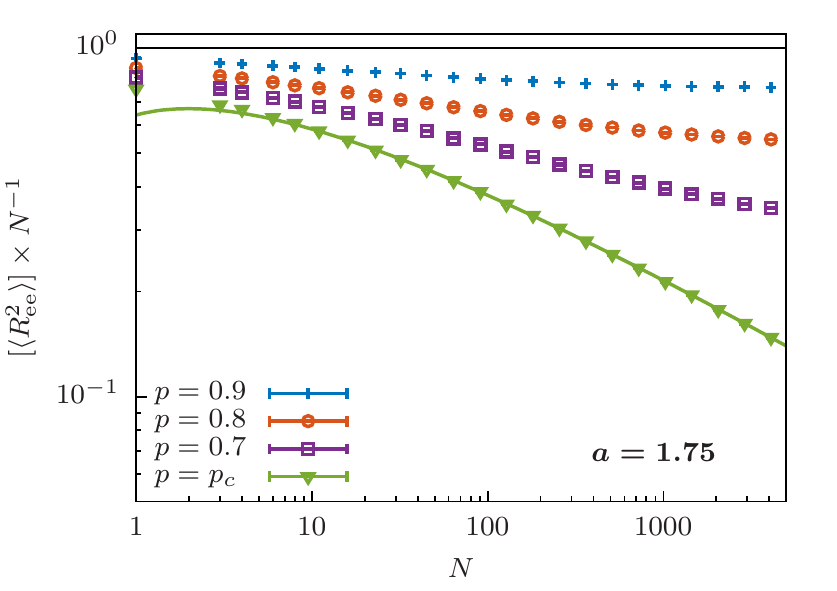}\\
  \includegraphics[width=0.43\textwidth]{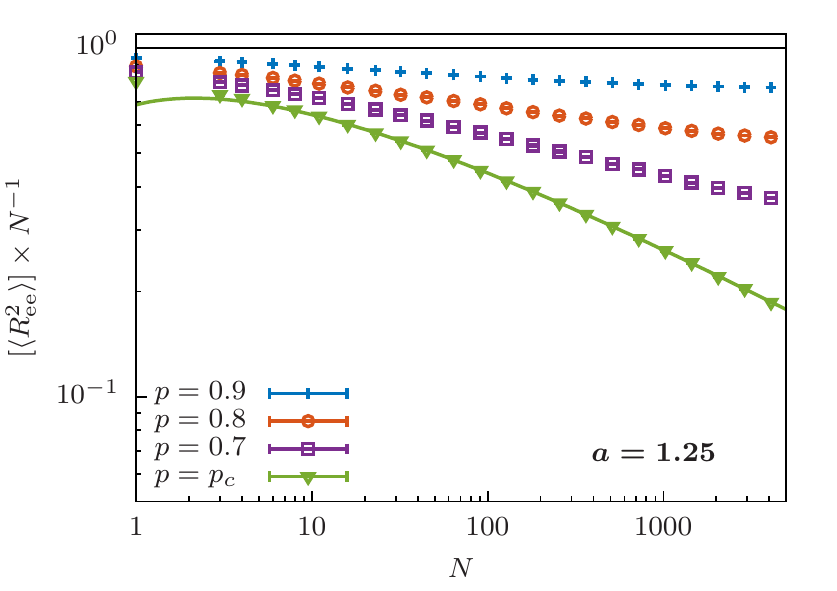}
  \includegraphics[width=0.43\textwidth]{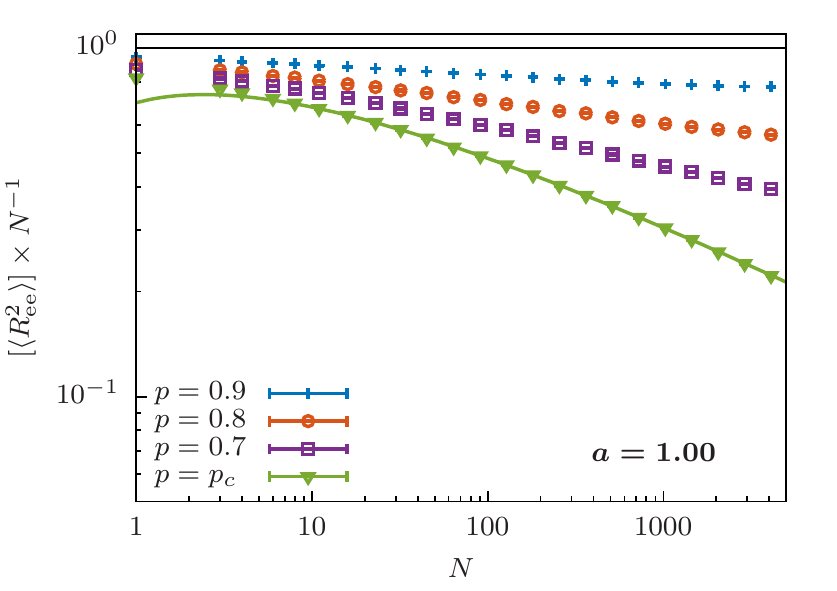}\\
  \includegraphics[width=0.43\textwidth]{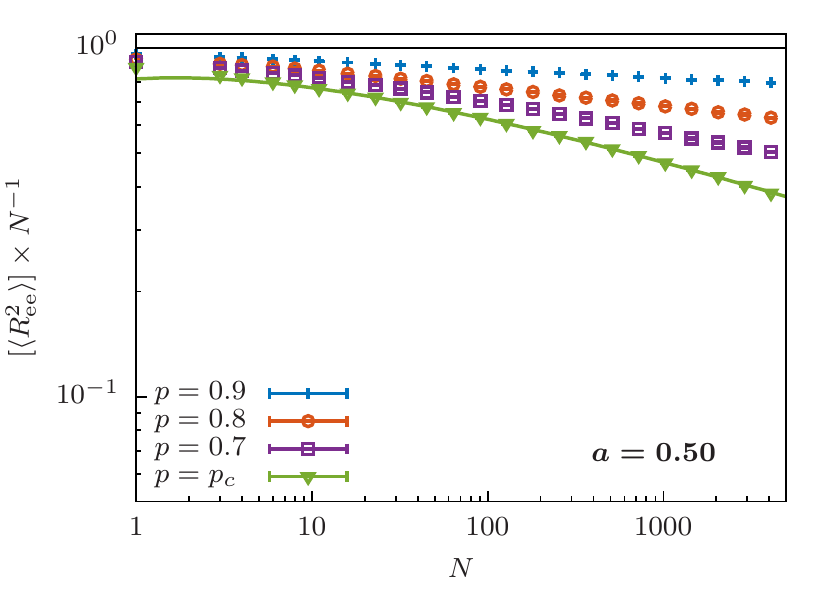}
  \includegraphics[width=0.43\textwidth]{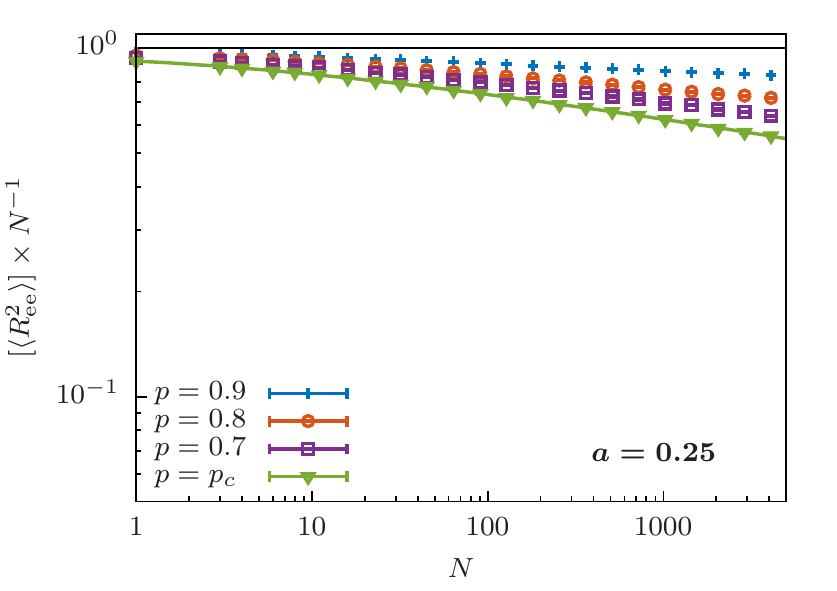}
  \caption{\label{figRWs} (Color online) Quenched disorder averages of the mean squared end-to-end distance
  per step for random walks on incipient percolation clusters
  for several degrees of correlation increasing from the top left to the bottom
  right panel and varying concentrations. The horizontal lines correspond to
  normal diffusion and the (green) lines for $p = p_{\text c}$ are least-squares fits of
equation~\eqref{nu_fit}.}
\vspace{2ex}
\includegraphics[width=0.43\textwidth]{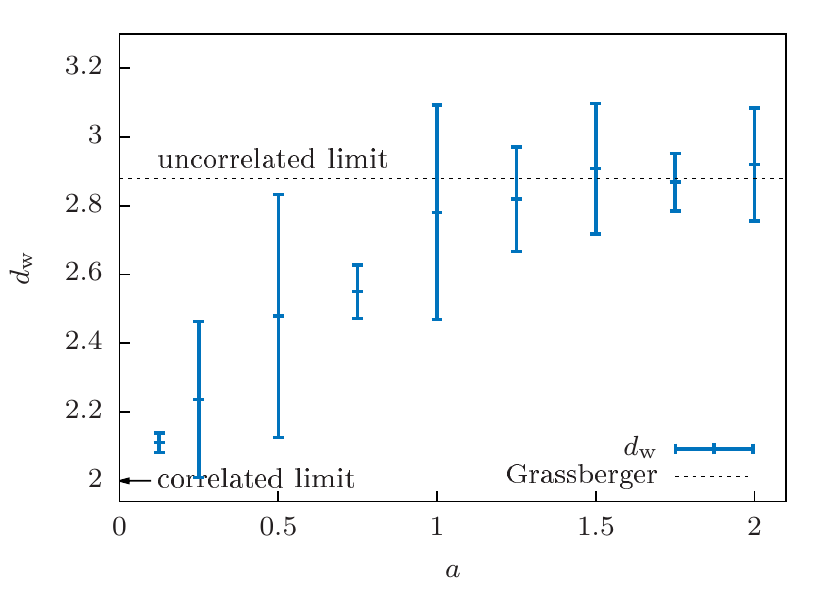}
  \caption{\label{figRWdw} (Color online)
    Random walk dimension $d_{\text w}$ at the percolation threshold $p_{\text c}$ as a
    function of correlation exponent $a$. The horizontal dotted line shows the
    limiting value for the uncorrelated case when
    $a\rightarrow\infty$~\cite{Grassberger1999}.
  }
\end{figure}

We performed independent averages for walks of different maximal lengths $N_i$,
which were chosen as rounded powers of $\sqrt{2}$. For each length, we only
analyzed the final displacement, thus obtaining completely uncorrelated
estimates for the average displacement after each length $N_i=2^{i/2}$, $0\leqslant i
\leqslant 24$. The results for varying
correlation exponents $a$ and
levels of concentration $p$
are shown in figure~\ref{figRWs}. For the uncorrelated case, shown on
the top left, the behavior is as expected: for $p>p_{\text c}$, the curves approach
horizontal lines corresponding to normal diffusion, while the
asymptotic behavior is characterized by a smaller exponent for $p=p_{\text c}$. This also seems to
be the case for $a=1.75$ (top right of figure~\ref{figRWs}), where we are still
above the Harris boundary of $a=3/2$. For stronger correlations, it is hard to
tell from these data whether or not the walks for $p>p_{\text c}$ asymptotically follow
the normal diffusive behavior. The curves for $a=1.25$
and $a=1.0$ seem to level off slightly, but the systems are too small to clearly reflect the asymptotic behavior.
We did not find a convincing fitting approach for the data above $p_{\text c}$,
and the question whether long-range correlations can alter the walks' asymptotic
scaling behavior even above criticality thus remains unsolved. The values at the
critical concentration, on the other hand, are nicely described by power-law fits with confluent correction term:
\begin{equation}\label{nu_fit}
	\left [ \left \langle R^2 \right \rangle \right ]_{\text{av}}= AN^{2\nu_{\text{RW}}}\left(1+bN^{-\Delta}\right ).
\end{equation}
Note that for uncorrelated percolation, the exponent of the correction term is
known very accurately~\cite{Kammerer2008, Ziff2011},
$\Delta = 3/2 d_{\text w} = 0.5211(2)$.

The resulting estimates for the walk dimensions $d_{\text w} = 1/\nu_{\text{RW}}$ at criticality are plotted in
figure~\ref{figRWdw} and listed in table~\ref{tab_exponents}. As can be seen, they
are consistent with the estimate for uncorrelated percolation for $a\geqslant 1.0$.
	For stronger correlations, the values decrease smoothly towards the full-lattice value of
	$d_{\text w}=2$ \linebreak for normal diffusion. This limit behavior is intuitively clear since any infinitely strongly
correlated lattice will almost always either be fully occupied or empty.
The agreement of $d_{\text w}$ with the uncorrelated estimate for large $a$ is
consistent with the theoretical predictions by Weinrib and
Halperin~\cite{Weinrib1984} that above the extended Harris threshold $a=3/2$, the system
effectively behaves as uncorrelated. However, the uncorrelated behavior of $d_{\text w}$ persists
even below the threshold down to $a \approx 1$.
This is in agreement with observations for the fractal
dimension, which also keeps its uncorrelated value well below
the Harris threshold~\cite{Schrenk2013}.

\begin{table}[!t]
	\caption{Results for the percolation threshold $p_{\text c}$
and the walk dimension $d_{\text w}$ for different correlation exponents $a$ at
the percolation threshold $p_{\text c}$. The errors provided refer to the fit errors.\vspace{2ex}
}\label{tab_exponents}
\begin{center}
\begin{tabular}{ |l |l| l|}\hline \hline
	$a$ & $p_{\text c}$ &  $d_{\text w}$ \\
  \hline\hline
  $\infty$ & $0.592746$~{\cite{Newman2001,Ziff1994}} &  $ 2.8784(8)$~\cite{Grassberger1999}\\
           &                              &  2.92(5) \\\hline
  3        & $0.55214(1)$                 &
                                               2.86(5) \\\hline
  2.5      & $0.54790(1)$                 &
					       2.92(11) \\\hline
  2        & $0.54299(2)$                 &
					       2.92(17) \\\hline
  1.75     & $0.54018(2)$                 &
					       2.87(9) \\\hline
  1.5      & $0.53710(2)$                 &
					       2.91(19) \\\hline
  1.25     & $0.5337(1)$                  &
                                               2.82(16) \\\hline
  1        & $0.5300(1)$                  &
					       2.78(32) \\\hline
  0.75     & $0.5255(2)$                  &
					       2.55(8) \\\hline
  0.5      & $0.5209(4)$                  &
					       2.48(36) \\\hline
  0.25     & $0.514(2)$                   &
					       2.24(23) \\\hline
  0.125    & $0.506(4)$                   &
					       2.11(3) \\
  \hline \hline
\end{tabular}
\end{center}
\vspace{-3mm}
\end{table}

\section{Conclusions and prospects}\label{sect_summary}

We investigated percolation on a square lattice with long-range
correlated disorder. The correlations were generated using the improved Fourier
filtering method (FFM) and carefully verified to decay according to the desired
power law. The percolation threshold has been measured for different values of
the correlation exponent $a$. The results strongly depend on the details of the
method, cf.\ \cite{Prakash1992}. Here, we used the FFM as proposed by Makse et
al.~\cite{Makse1996} combining an analytic continuum solution of the
spectral density in infinite space with a discrete Fourier transform on a
finite lattice. While this works well to the leading order, it introduces some
subtle difficulties such as the adjustment of the zero mode
or a variance of the
desired distribution that deviates from unity.
While this affects non-universal quantities such as the
percolation threshold or the amplitudes in scaling laws, universal
exponents such as
the walk dimension~$d_{\text w}$ are not
sensitive to such details.

We then performed an exact enumeration of random walks to investigate the properties of
correlated percolation clusters as a medium for diffusion. The walk dimension $d_{\text w}$ on critical
clusters was determined as a function of the correlation exponent $a$. For weak
correlations (large $a$), the uncorrelated behavior is recovered, and this behavior even seems to
prevail below the threshold set by the extended Harris criterion~\eqref{Harris}. For increasing correlations (small
$a$), $d_{\text w}$ was found to approach the normal diffusion value of $2$.
The results on super-critical clusters ($p>p_{\text c}$) are quantitatively less
conclusive since the correlations strongly aggravate the finite-size effects. However,
for the uncorrelated and the weakly correlated case, we qualitatively recovered
the expected behavior of normal diffusion. In the strongly correlated case,
persistence of anomalous diffusion cannot be ruled out from our data.

It would be very interesting to also study the behavior of self-avoiding walks
on long-range correlated percolating clusters. For the uncorrelated case, there
exist exact enumeration techniques that enable treatment of extremely long
walks~\cite{Fricke2012,Fricke2014}. While this approach becomes more demanding for an
increasing correlation strength ($a\rightarrow0$), it should still be applicable to some extent.
It could be supplemented by chain-growth Monte Carlo
approaches such as PERM~\cite{Grassberger1997}, which in the past were also used to study the uncorrelated system~\cite{Blavatska2008,Blavatska2008a}.
With some efforts, this would enable one to compare with
field-theoretic predictions for
polymers~\cite{Blavatska2001, Blavatska2001b, Blavatska2002, Blavatska2005}.

%\vspace{-2mm}
\section*{Acknowledgements}

The article is dedicated to Professor Yurko Holovatch on the occasion of his
60th birthday.
The cooperation with the Lviv group was supported by an
Institute Partnership Grant of the Alexander von Humboldt Foundation (AvH)
under Grant No.~3.4--Fokoop--DEU/1117877,
the Deutsch-Franz\"osische Hochschule (DFH) through the
International Doctoral College
``${\mathbb L}^4$'' Leipzig-Lorraine-Lviv-Coventry under
Grant \linebreak No.~CDFA-02-07, and by the
EU Marie Curie IRSES Network DIONICOS under
Contract No.~PIRSES-GA-2013-612\,707.

The work was funded by the Deutsche Forschungsgemeinschaft (DFG) via
FOR~877 (project P9) under Grant No.~JA~483/29-1 and
Sonderforschungsbereich/Transregio SFB/TRR~102 (project~B04).
We are grateful for further support from the
Leipzig Graduate School ``BuildMoNa''--Building with Molecules and
Nano-objects. JZ received financial support from the
 German Ministry for Education and Research (BMBF) via the
 Bernstein Center for Computational Neuroscience (BCCN)
 G{\"o}ttingen under Grant \linebreak No.~01GQ1005B.

%\vspace{-2mm}

\ukrainianpart
%\vspace{-2mm}
\title{Закони скейлінгу для випадкових блукань у далекосяжно-скорельованих невпорядкованих середовищах}
\author{Н. Фріке\refaddr{label1}, Й. Ціренберг\refaddr{label1,label2,label3}, М. Маренц\refaddr{label1}, Ф.П. Шпіцнер\refaddr{label1},
 В. Блавацька\refaddr{label4}, В. Янке\refaddr{label1} }
\addresses{
\addr{label1} Інститут теоретичної фізики, Університет Ляйпціґа, D-04009 Ляйпціґ, Німеччина
\addr{label2} Центр обчислювальної нейробіології ім. Бернштейна, D-37077 Ґеттінґен, Німеччина
\addr{label3} Інститут динаміки і самоорганізації Макса Планка, D-37077 Ґеттінґен, Німеччина 
\addr{label4} Інститут фізики конденсованих систем НАН України,
вул. Свєнціцького, 1, 79011 Львів, Україна
}

\newpage
\makeukrtitle
\begin{abstract}
\tolerance=3000%

Ми досліджуємо закони скейлінгу для дифузії у двовимірному середовищі із далекосяжно-скорельованим безладом шляхом точного підрахунку випадкових блукань.
Невпорядковане середовище моделюється як перколяційний кластер із кореляціями, що спадають з відстанню згідно степеневого закону $r^{-a}$,
згенерований за допомогою покращеного методу фільтрування Фур'є. Щоб охарактеризувати такий тип безладу, визначаємо поріг перколяції $p_{\text c}$
шляхом дослідження імовірностей появи безмежного кластера. При $p_{\text c}$  ми оцінюємо вимірність (суб-дифузивного) блукання $d_{\text w}$ при різних значеннях кореляційного показника~$a$. Вище $p_{\text c}$ наші результати вказують на поведінку звичайних випадкових блукань при слабких кореляціях, в той час як не можна виключити аномальну дифузію у випадку сильних кореляцій, тобто при малих~$a$.
\keywords далекосяжно-скорельований безлад, критичні перколяційні кластери, випадкові блукання, точний підрахунок, закони скейлінгу

\end{abstract}

\end{document}